\documentclass[10pt,onecolumn]{IEEEtran}
\usepackage{amssymb}
\usepackage{bbm}
\ifCLASSINFOpdf
  \usepackage[pdftex]{graphicx}
 \DeclareGraphicsExtensions{.pdf,.jpeg,.png}
\else
\fi
\usepackage[cmex10]{amsmath}
\usepackage{tikz}
\usepackage{tikz}
\usepackage{bigdelim}
\usepackage{graphicx}
\usepackage{epstopdf}
\DeclareGraphicsExtensions{.eps}

\newtheorem{thm}{Theorem}
\newtheorem{lem}{Lemma}

\newenvironment{definition}[1][Definition]{\begin{trivlist}
\item[\hskip \labelsep {\bfseries #1}]}{\end{trivlist}}

\newcommand{\qed}{\nobreak \ifvmode \relax \else
      \ifdim\lastskip<1.5em \hskip-\lastskip
      \hskip1.5em plus0em minus0.5em \fi \nobreak
      \vrule height0.75em width0.5em depth0.25em\fi}
\begin{document}
\baselineskip 8.0 mm
\title{Degrees of Freedom Rate Region of the $K$-user Interference Channel with Blind CSIT Using Staggered Antenna Switching}
\author{\begin{normalsize}Milad Johnny and Mohammad Reza Aref
\\Information System and Security Lab (ISSL),\\Sharif Universiy of Technology, Tehran, Iran\\E-mail: Johnny@ee.sharif.edu, Aref@sharif.edu
\end{normalsize}
}
\maketitle
\begin{abstract}
In this paper, we consider the problem of the interference alignment for the $K$-user SISO interference channel with blind channel state information at transmitters (CSIT). Our achievement in contrast to popular $K-$user interference alignment (IA) scheme has more practical notions. In this case every receiver is equipped with one reconfigurable antenna which tries to place its desired signal in a subspace which is linearly independent from interference signals. We show that if the channel values are known to the receivers only, the sum degrees-of-freedom (DOF) rate region of the linear BIA with staggered antenna switching is $\frac{Kr}{r^2-r+K}$, where $r =  \left \lceil{\frac{\sqrt{1+4K}-1}{2}} \right \rceil$. The result indicates that the optimum DoF rate region of the $K-$user interference channel is to achieve the DoF of $\frac{\sqrt{K}}{2}$ for an asymptotically large network. Thus, the DoF of the $K$-user interference channel using staggered antenna switching grows sub-linearly with the number of the users, whereas it grows linearly in the case where transmitters access the CSI. In addition we propose both achievability and converse proof so as to show that this is the DoF rate region of blind interference alignment (BIA) with staggered antenna switching.
\end{abstract}


\IEEEpeerreviewmaketitle
\section{Introduction}
\IEEEPARstart{T}{he} new increasing demand for higher data rate communication motivates researchers to introduce new tools to reduce channel constrains such as interference in the transmission medium. In the network area, due to high speed of progressing, the opportunities for innovation and creativity increases. Interference channel due to its important role in today's communication systems has been the focus of attention in today's wireless networks. The importance of the problem of finding the capacity of interference channel is so essential that after point-to-point communication scenario it is the second problem which was introduced by Shannon \cite{shannon} and it has many applications in today's communication networks. Unfortunately finding the exact capacity of the interference channel is so hard that it is still open for near half of the century. While finding the exact capacity of many networks is still open, DoF can analyze capacity characteristics of such networks at high $\mathrm{SNR}$ regions. Recently \cite{2}, by the basic idea of IA with some constraints shows that one can achieve $\frac{K}{2}$ DoF for the fast fade interference channel. This method for practical cases where transmitters do not have access to channel values fail to get any achievement. The CSI was not the only barrier for implementation of such a method, the long precoder size at transmitters and the high speed channel changing pattern show further impractical aspects of such a method. Such assumption is hard to materialize under any practical channel feedback scheme. To combat the CSIT problem, there are two different strategies which are related to blind CSIT and outdated CSIT (delay CSIT). Moreover, as a first step to study the impact of the lack of channel knowledge, \cite{milad} shows that with some conditions on the direct and interference channels, one can perfectly or imperfectly align interference; if half of the interference channel values are not available at both the transmitters and receivers, one can achieve the DoF of $\frac{K}{2}$. In \cite{Gou}, the authors show that artificially manipulating the channel itself to create the opportunities, one can facilitate BIA. They equip each user with simple staggered antenna which can switch between multi-mode reception paths. In this work by the use of staggered antenna switching one can achieve $\frac{MK}{M+K-1}$ DoF for the well-known MISO broadcast channel where each receiver is equipped with multi-mode antenna. After finding DoF rate region of MISO broadcast channel in the case of delay CSIT \cite{mad} there are several works characterizing the DoF of the interference channel with the delayed CSIT. In \cite{8}, with the assumption of delay CSIT it is shown that the DoF of the $K$-user interference channel can achieve the value of $4/(6 \, \ln(2)  - 1) \approx 1.266$ as $K\rightarrow \infty$. The BIA scheme with staggered antenna switching only requires multi-mode antenna switching at the receivers, which does not need any significant hardware complexity \cite{4}. In \cite{7}, for the 3-user interference channel Wang shows that using staggered antenna switching one can achieve the sum DoF of $\frac{6}{5}$ in the case of blind CSIT. Alaa and Ismail in \cite{Alaa}, trie to generalize the DoF rate region of 3-user interference channel with staggered antenna switching to the $K-$user interference channel but to some extend it has contradiction with our work for the $K>6$ and evident case of $K=1$ where DoF is one. In this paper, we generalize Wang problem for the case of $K-$user interference channel and we show that with the aid of multi-mode antenna switching at receivers, the sum DoF is $\max_{r}{\frac{Kr}{r^2-r+K}}$. This result indicates that when the number of the users $K$ limits to infinity, BIA can achieve $\frac{\sqrt{K}}{2}$ DoF which is in contradiction with the DoF upper-bound of $\frac{2K}{K+2}$, thus the sum DoF does not scale linearly with $K$ as in the case when CSIT is available, but rather scales {\it sub-linearly} with the number of users.
\subsection{Organization}
This paper is organized as follows. The next section describes the system model. In Section III, we explore overviews of the main result. In section IV, by providing both achievability and converse proofs we show that $\max_{r}{\frac{Kr}{r^2-r+K}}$ is the sum DoF rate region of the $K-$user interference channel with staggered antenna switching. Finally, we draw our conclusions in Section V.

\section{System Model}
\begin{figure}
  \centering
  \includegraphics[width=0.5\textwidth]%
    {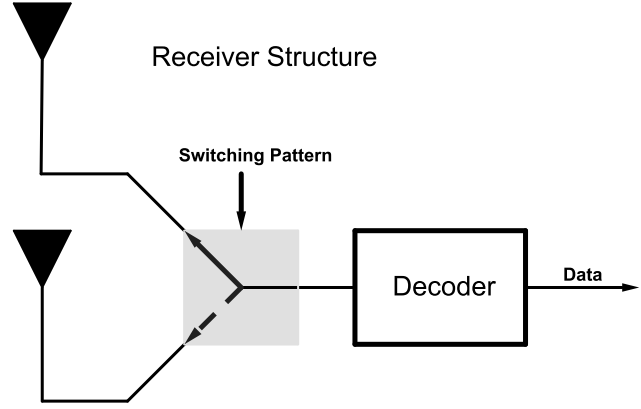}% picture filename
  \caption{Structure of the two-mode staggered antenna switching. In this case every receiver equipped with two antennas and a switch which can select between two different modes.}
\end{figure}
Consider the $K-$user interference channel, where each receiver has more than one receiving antenna. In this case at each time snapshot all the receivers can switch to one of the receiving antennas to receive its desired signal from corresponding transmitter and all other transmitters as interferences (see Figure 1). This channel consists of $K$ transmitters ${\left \{\mathrm{TX}_k  \right \}_{k=1}^{K}}$ and $K$ receivers ${\left \{\mathrm{RX}_k  \right \}_{k=1}^{K}}$. Let a discrete interference channel be $ K^{2}+2K$ tuple $\left({\bf \bar{H}}^{[11]},{\bf \bar{H}}^{[12]},...,{\bf \bar{H}}^{[KK]},{\bf \bar{{x}}}^{[1]},...,{\bf \bar{{x} }}^{[K]},{\bf \bar{{y}}}^{[1]},...,{\bf \bar{{y}}}^{[K]}\right)$, where $\left(\bf \bar{{x}}^{[1]},...,\bf \bar{{x}}^{[K]}\right)$
and $\left({\bf \bar{{y}}}^{[1]},...,{\bf \bar{{y}}}^{[K]}\right)$ are $K$ finite input and output of the channel respectively; in the interference channel, the input of $k^{th}$ transmitter is represented by ${\bf \bar{{x}}^{[k]}}=[{x}_{1}^{[k]},....,{x}_{n}^{[k]}]^T$. Similarly the output of the channel can be represented by column matrix of ${\bf \bar{{y}}}^{[k]}=[{{y}_{1}}^{[k]},....,{{y}_{n}}^{[k]}]^T$.
For a specific case where the thermal noise power is zero, ${\bf \bar{H}}^{[pq]}$ is a collection of such a diagonal matrices which maps ${\bf \bar{{x}}}^{[q]}$ to received signal at $p^{th}$ receiver and represents channel model.
 Therefore, the received signal at the $k^{th}$ receiver consisting of $n$ time snapshot channel uses can be represented as follows:
\begin{equation}
\label{1}
{\bar{{\bf y}}^{[p]}} = \sum_{q=1}^{K} {{\bf H}^{[pq]}} {{\bf \bar{x}}^{[q]}} + {{\bf \bar{z}}^{[p]}}, \,\, p, q \in \{1,2,...,K\}
\end{equation}
where ${{\bar{\bf y}}^{[p]}}$ represents the received signal over $n$ channel uses (time or frequency slots), ${{\bar{\bf x}}^{[q]}}$ is the transmitted signal vector by the $q^{th}$ transmitter subject to average power constraint of ${\lim_{n \rightarrow \infty}{\frac{1}{n} \sum_{i=1}^{n}{\left({{x_i}^{[k]}}\right)^2}}}=\mathrm{SNR}$, ${\bar{ \bf{z}}}^{[p]}$ is an additive white Gaussian noise in which $\lim_{n \rightarrow \infty}{\frac{1}{n}\mathrm{tr}\left({{\bar {\bf{z}}}^{[p]}}{\left({\bar{\bf{z}}}^{[p]}\right)}^{H}\right)}=1 $, and ${{\bar{\bf H}}^{[pq]}}$ is a diagonal matrix representing the channel model between the $\mathrm{TX}_q$ and $\mathrm{RX}_p$. The channel matrix can be written as:
\begin{equation}
\label{2}
{{\bar {\bf{H}}}^{[pq]}} = \mathrm{diag}\left( \left[ {h^{[pq]}_1,h^{[pq]}_2,\dots,h^{[pq]}_n} \right]\right),
\end{equation}
where depending on the number of antenna modes $h^{[pq]}_j \in \lbrace{h^{[pq]}(1),h^{[pq]}(2),\dots,h^{[pq]}(M)}\rbrace$. In other words the diagonal matrix $\bar{\bf{H}}^{[pq]}$ can be represented as follows:
\begin{equation}
{{\bf \bar{H}}^{[pq]}} = \mathrm{diag}([h^{[pq]}\left({\bf{SW}}_p {(1)}\right) \,\, h^{[pq]}\left({\bf{SW}}_p {(2)}\right) \,\, ... \,\, h^{[pq]}\left({\bf{SW}}_p {(n)}\right)]),
\end{equation}
where ${\bf{SW}}_p = \left[ {\bf{SW}}_p {(1)}~{\bf{SW}}_p {(2)}~ \dots~{\bf{SW}}_p {(n)} \right] $,  ${\bf{SW}}_p {(j)} \in \{1,\dots,M\}$ shows the switching pattern matrix at $\mathrm{RX}_p$. This switching pattern for all channels which end in the same destination e.g. $p$ have the same effect. We assume all the channel links between different transceivers are constant for $n$ channel uses. Also, ${{\bf \bar{x}}^{[q]}}$ is a column matrix with the size of $n \times 1$ and can be represented as follows:
\begin{equation}
\label{3}
{{\bf \bar{x}}^{[q]}} = \sum_{d=1}^{d_{q}} x_{d}^{[q]} \, {{{\bf v}_{d}}^{[q]}}
\end{equation}
where $d_{q}$ is the number of symbols transmitted by the $q^{th}$ user over $n$ channel uses, $s_{d}^{[q]}$ is the $d^{th}$ transmitted symbol and ${{{\bf v}_{d}}^{[q]}}$ is an $n \times 1$ transmit beamforming vector for the $d^{th}$ symbol. The equation of \eqref{3} can be defined and simplified as follows:
\begin{equation}
{{\bf \bar{x}}^{[q]}}= \bar{\bf{V}}^{[q]} \bf{X}^{[q]},
\end{equation}
where ${\bf{X}}^{[q]}=\left[ x_{1}^{[q]},\dots,x_{d_q}^{[q]}\right]^{T}$ and ${\bar{\bf V}^{[q]}} = [{\bf v}_{1}^{[q]} \,\, {\bf v}_{2}^{[q]} \,\, ...\,\, {\bf v}_{d_q}^{[q]}]$. Also, ${\bf v}_{d}^{[q]}$ is one of the basic vectors of designed precoder at $\mathrm{TX}_q$.
\subsection{Degrees of Freedom for the $K-$user Interference Channel}
In the $K$-user BIA interference channel using staggered antenna switching, we define the degrees of freedom region as follows\cite{1}:
\begin{align}
\Bigg\{(d_1,d_2,\dots,d_K)\in \mathbb{R}_{+}^{K}:&\forall (w_1,\dots,w_K) \in \mathbb{R}_{+}^{K},& \\
&{w_1}{d_1}+\dots + {w_K}{d_K}\nonumber\leq\lim_{\rho \to \infty} \sup\left [\underset{\mathcal{\emph{R}(\rho) \in \mathcal{C}(\rho) }}{\sup}\frac {(w_1{R}_{1}(\rho)+\dots+w_K{R}_{K}(\rho))}{\log(\rho)}\right ]&\Bigg\}.
\end{align}
\section{Overview of the Main Result}
In this paper we explore interference alignment for the $K-$user interference channel with blind CSIT. We provide both achievability and the DoF upper bound by the linear interference alignment. The summary of the results can be expressed by the following theorem.
\begin{thm}
{\it The number of DoFs for the $K$-user SISO interference channel with BIA using staggered antenna switching is $\max_{r}{\frac{Kr}{r^2-r+K}}, r\in \mathbb{N}$.}
\end{thm}
 The result indicates that when the number of users limits to infinity and there is not any information at transmitters about CSI, the number of DoFs goes to $\frac{\sqrt{K}}{2}$.   

\section{Outer Bound on the Degrees of Freedom for the BIA $K-$user Interference Channel Using staggered Antenna Switching}
In this section, we derive an upper bound on the sum DoF of the interference channel with BIA using staggered antenna switching at the receivers. In the next theorem, we assume no CSIT, each receiver is equipped with a reconfigurable antenna with an arbitrary number of antenna modes, and each transmitter has a conventional antenna.
\begin{figure}
  \centering
  \begin{tikzpicture}
  \draw (-12,4) ellipse (0.8cm and 1.5cm);
  \draw (-9,4) ellipse (0.8cm and 1.5cm);
  \draw (-12,6) circle (0.05cm);
  \node            (a) at (-12.25,6){$1$};
  \draw (-12,5) circle (0.05cm);
  \node            (b) at (-12.25,5){$l_1$};
  \draw (-12,4.5) circle (0.05cm);
  \node            (c) at (-12.25,4.5){$q_1$};
  \node            (c) at (-12.25,4){$q_2$};
  \draw (-12,4) circle (0.05cm);
  \node            (d) at (-12,3.75){$\vdots$};
  \draw (-12,3) circle (0.05cm);
  \node            (e) at (-12.25,3){$l_r$};
  \draw (-12,2) circle (0.05cm);
  \node            (f) at (-12.25,2){$q_4$};
  \draw (-12,1) circle (0.05cm);
  \node            (f) at (-12.25,1){$K$};
  \draw (-9,6) circle (0.05cm);
  \node            (g) at (-8.75,6){$1$};
  \draw (-9,5) circle (0.05cm);
  \node            (h) at (-8.75,5){$l_1$};
  \draw (-9,4.5) circle (0.05cm);
  \node            (i) at (-8.75,4.5){$q_3$};
  \draw (-9,4) circle (0.05cm);
  \node            (j) at (-9,3.75){$\vdots$};
  \draw (-9,3) circle (0.05cm);
  \node            (k) at (-8.75,3){$l_r$};
  \draw (-9,2) circle (0.05cm);
  \node            (l) at (-8.75,2){$q_4$};
  \draw (-9,1) circle (0.05cm);
  \node            (l) at (-8.75,1){$K$};
  \node            (m) at (-7,4){$l^{t}$ RX set};
  \node            (n) at (-14,4){$l^{t}$ TX set};
  \node            (o) at (-12.25,0){\begin{LARGE}
  TX
  \end{LARGE}};
  \node            (p) at (-9,0){\begin{LARGE}
  RX
  \end{LARGE}};
  \node            (tk) at (-11.75,1){};
  \node            (rk) at (-9.25,1){};
  \node            (t1) at (-11.75,2){};
  \node            (tlq) at (-11.75,4){};
  \node            (tq4) at (-11.75,6){};
  \node            (r1) at (-9.25,2){};
  \node            (rlq) at (-9.25,4){};
  \node            (rq4) at (-9.25,6){};
  \draw[->] (t1) edge (r1);
  \draw[->] (t1) edge (rlq);
  \draw[->] (t1) edge (rq4);
  \draw[->] (tlq) edge (r1);
  \draw[->] (tlq) edge (rlq);
  \draw[->] (tlq) edge (rq4);
  \draw[->] (tq4) edge (r1);
  \draw[->] (tq4) edge (rlq);
  \draw[->] (tq4) edge (rq4);
  \draw[->] (t1) edge (rk);
  \draw[->] (tk) edge (r1);
  \draw[->] (tk) edge (rk);    
  \end{tikzpicture}
  \caption{In this figure we show transceivers number of the set $l^{t}=\{l_1,l_2,\dots,l_r\}$ with the closed circular shape. The complimentary transceivers out of this circular shape can be modeled by the set $\{1,\dots,K\}-l^{t}$. Also there is a connection between all transmitters and receivers but to avoid being so crowded we show a few of them.}
\end{figure}
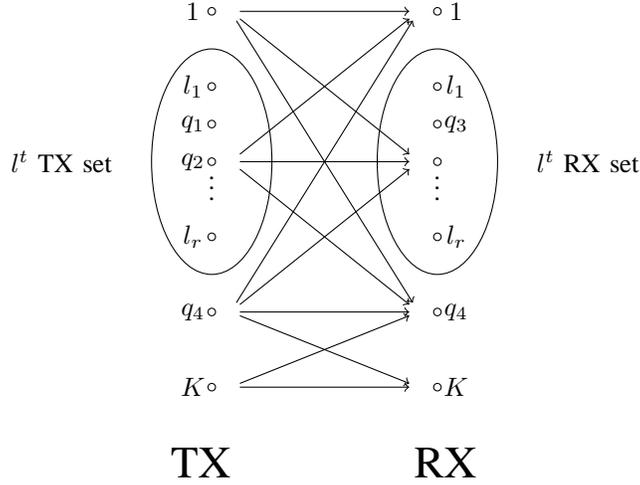
 Now consider the set $l^{t}=\{l_1,\dots,l_r\}\subseteq \{1,\dots,K\}$ where $\vert l^{t} \vert = r$ and $1 \leq t \leq \binom{K}{r}$. We assume every basic vector from each transmitter aligns with interference generated from $r-1$ transmitters at $K-r$ receivers. In other words, if ${\bf{v}}^{[q]}$ is one of the basic vectors of $q^{th}$ transmitter, we have:
\begin{equation}
\bar{{\bf{H}}}^{[pq]}{\bf{v}}^{[q]} \prec \bar{\bf{H}}^{[pq']}\bar{{\bf{V}}}^{[q']}.
\end{equation}
Where, $(q, q' \in l^{t},~q \neq q'$ and $p \in \{1,\dots,K\}-l^{t}$.\\
\textit{{Remark:}} $\bar{{\bf{H}}}^{[qq]}{\bf{v}}^{[q]}\notin \text{span} \left( \bar{{\bf{H}}}^{[qq']}\bar{{\bf{V}}}^{[q']} \right)$ if not, the desired signal space is polluted by interference.\\
\textit{{Lemma 1:}}
If ${\bf{v}}^{[q]}$ is aligned with interference of $r-1$ transmitters (in the set of $l^{t}$) at the members of the set $\{1,\dots,K\}-l^{t}$ receivers, it can not be aligned with the interference generated from the transmitters set of $l^{t}$ at $r-1$ receivers of the set $l^{t}-\{q\}$.\\
\textit{\textbf{Proof:}} 
Suppose that $q_1$ and $q_2$ are transmitters in the set of $l_t$. Also, $\mathrm{RX}_{q_3}$ is a receiver in the set of $l_t$ and $\mathrm{RX}_{q_4}$ is a receiver in the complimentary set of $l_t$ (in the set of $\{1,\dots,K\}-l_t$). From the assumption of this lemma we can assume:
\begin{equation}
\bar{{\bf{H}}}^{[q_4q_1]}{\bf{v}}^{[q_1]} \in \text{span}\left({\bar{{\bf{H}}}^{[q_4 q_3]}\bar{{\bf{V}}}^{[q_3]}}\right).
\end{equation}
From Lemma 2 of \cite{7}, since $\bar{{\bf{H}}}^{[q_4 q_1]}$ and $\bar{{\bf{H}}}^{[q_4 q_3]}$ are diagonal and have the same changing pattern, ${\bf{v}}^{[q_1]} \in \text{span}\left({\bar{{\bf{V}}}^{[q_3]}}\right)$.\\
Suppose not. We take the negation of the given statement and suppose it to be true. Assume, to the contrary, that:
\begin{equation}
\left\{ \exists q_3 \in l_t: \text{span}\left( \bar{{\bf{H}}}^{[q_{3}q_{1}]}{\bf{v}}^{[q_1]}\right)\in \text{span}\left( \bar{{\bf{H}}}^{[q_{3}q_{2}]}{\bar{{\bf{V}}}}^{[q_2]}\right) \right\}.
\end{equation}   
From this assumption we have:
\begin{equation}
\text{span}\left({\bar{{\bf{H}}}}^{[q_3 q_1]}{\bf{v}}^{[q_1]}\right) \in \text{span} \left({\bar{{\bf{H}}}^{[q_3 q_3]}} \left(\bar{{\bf{H}}}^{[q_3 q_3]}\right)^{-1}\bar{{\bf{H}}}^{[q_3 q_2]}{\bar{{\bf{V}}}^{[q_2]}}\right).
\end{equation}
Since ${\bar{{\bf{H}}}^{[q_3 q_1]}}$ and ${\bar{{\bf{H}}}^{[q_3 q_3]}}$ have similar changing pattern, we get:
\begin{equation}
\text{span}\left({\bf{v}}^{[q_1]}\right) \in \text{span} \left(\left(\bar{{\bf{H}}}^{[q_3 q_3]}\right)^{-1}\bar{{\bf{H}}}^{[q_3 q_2]}{\bar{{\bf{V}}}^{[q_2]}}\right).
\end{equation}
Therefore, since ${\bf{v}}^{[q_1]} \in \text{span}\left({\bar{{\bf{V}}}^{[q_3]}}\right)$, we have:
\begin{equation}
\text{dim} \left( {\bar{{\bf{V}}}^{[q_{3}]}} \cap \left(\bar{{\bf{H}}}^{[q_{3} q_{3}]}\right)^{-1}\bar{{\bf{H}}}^{[q_{3} q_{2}]}{\bar{{\bf{V}}}^{[q_{2}]}} \right) >0,
\end{equation}
and finally we get:
\begin{equation}
\text{dim} \left( \bar{{\bf{H}}}^{[q_{3} q_{3}]}{\bar{{\bf{V}}}^{[q_{3}]}} \cap \bar{{\bf{H}}}^{[q_{3} q_{2}]}{\bar{{\bf{V}}}^{[q_{2}]}} \right) >0.
\end{equation}
The above relation shows that the desired signal $\bar{{\bf{H}}}^{[q_{3} q_{3}]}{\bar{{\bf{V}}}^{[q_{3}]}}$ at $q_{3}^{th}$ receiver has been polluted by the interference of $q_{2}^{th}$ transmitter. Hence by the assumption of $\left\{ \exists q_3 \in l_t: \text{span}\left( \bar{{\bf{H}}}^{[q_{3}q_{1}]}{\bf{v}}^{[q_1]}\right)\in \text{span}\left( \bar{{\bf{H}}}^{[q_{3}q_{2}]}{\bar{{\bf{V}}}}^{[q_2]}\right)\right\}$ we have a contradiction. This contradiction shows that the given assumption is false and the statement of the lemma is true. So, this completes the proof.\\
\begin{definition}
$d_{{i_1}{i_2}\dots{i_r}},~i_1 \neq i_2 \neq \dots \neq i_r$ shows the number of dimensions which is occupied by transmitters $i_1$,$i_2$,... and $i_r$ at $j^{th}$ receiver, where $j\notin \{i_1,i_2,\dots,i_r\}$. Also for every $i'_1,\dots,i'_r \in \{i_1,i_2,\dots,i_r\}$ we have:
\begin{equation}
d_{{i_1}{i_2}\dots{i_r}}=d_{{i'_1}{i'_2}\dots{i'_r}}.
\end{equation}
\end{definition}
\subsection{Converse Proof:}
The converse proof follows from the following upper bound on the DoF of the $K-$user interference channel with BIA.
At $j^{th}$ receiver the interference signal from transmitters $i_1$,$i_2$,... and $i_r$, where $j\notin \{i_1,i_2,\dots,i_r\}$ occupy $d_{{i_1}{i_2}\dots{i_r}}$ dimensions. In other words, every shared vectors between $r$ different users ($i_{1}^{th}$,$i_{2}^{th}$,... and $i_{r}^{th}$ users) occupy just only one dimension at $j^{th}$ receiver. On the other hand the total number of dimensions is $n$. Therefore, at $j^{th}$ receiver we have:
\begin{equation}
\label{14}
d_1+d_2+\dots+d_K-\left(r-1\right) \sum_{i_1,\dots,i_r}{d_{i_1,\dots,i_r}} \leq n,~i_1,\dots,i_r \in \{1,\dots,K\}-\{j\},
\end{equation}
where, the coefficient $(r-1)$ comes from this fact that $d_{i_1,\dots,i_r},~i_1,\dots,i_r \in \{1,\dots,K\}-\{j\}$ just only occupy one dimension at $j^{th}$ receiver while it counts $r$ times when we calculate $d_1+d_2+\dots+d_K$. Similarly at all the receivers we have:
\begin{equation}
\label{15}
\begin{aligned}
&\mathrm{RX}_{1}:~d_1+d_2+\dots+d_K-\left(r-1\right) \sum_{i_1,\dots,i_r}{d_{i_1,\dots,i_r}} \leq n,~i_1,\dots,i_r \in \{1,\dots,K\}-\{1\}&\\
&\mathrm{RX}_{2}:~d_2+d_1+\dots+d_K-\left(r-1\right) \sum_{i_1,\dots,i_r}{d_{i_1,\dots,i_r}} \leq n,~i_1,\dots,i_r \in \{1,\dots,K\}-\{2\}&\\
&\vdots&\\
&\mathrm{RX}_{K}:~d_K+d_1+\dots+d_{K-1}-\left(r-1\right) \sum_{i_1,\dots,i_r}{d_{i_1,\dots,i_r}} \leq n,~i_1,\dots,i_r \in \{1,\dots,K\}-\{K\}.&\\
\end{aligned}
\end{equation}
Adding all the above relations we conclude that:
\begin{equation}
\label{DOFlower}
K \sum_{i=1}^{K}{d_i}+\left(K-1\right)\left(r-1\right)\sum_{i_1,\dots,i_r}{d_{i_1,\dots,i_r}} \leq Kn,
\end{equation} 
in addition, it is clear that:
\begin{equation} 
r \sum_{i_1,\dots,i_r}{d_{i_1,\dots,i_r}} \geq \sum_{i=1}^{K}{d_i}.
\end{equation}
Since for $r \geq 1$ the value of $\left( K-1 \right) \geq \left(K-r\right)$ we have:
\begin{equation} 
\left( K-1 \right) r \sum_{i_1,\dots,i_r}{d_{i_1,\dots,i_r}} \geq \left(K-r\right) \sum_{i=1}^{K}{d_i},
\end{equation}
which shows that:
\begin{equation} 
\sum_{i_1,\dots,i_r}{d_{i_1,\dots,i_r}} \geq \frac{\left(K-r\right)}{\left( K-1 \right) r } \sum_{i=1}^{K}{d_i}.
\end{equation}
Therefore from \eqref{DOFlower} we have:
\begin{equation}
\label{DOFlower2}
K \sum_{i=1}^{K}{d_i}+\left(K-1\right)\left(r-1\right)\frac{\left(K-r\right)}{\left( K-1 \right) r } \sum_{i=1}^{K}{d_i} \leq Kn.
\end{equation}
After simplifying \eqref{DOFlower2} we get:
\begin{equation}
\frac{\sum_{i=1}^{K}{d_i}}{n} \leq \frac{Kr}{r^2-r+K} \leq max_{r}{\frac{Kr}{r^2-r+K}},
\end{equation}
thus, we complete the converse proof.\\
In order to find the maximum value of $d(r)$ we analyze the continuous function of $f(x)=\frac{Kx}{x^2-x+K}$. The first derivation of this function has just one positive root of $x=\sqrt{K}$ which shows that it has just only one extremum point. Also it can easily be shown that for $x\geq 0$ the function $f(x)$ is greater than or equal to zero. Since $f(x=0)=0$ and $f(x\rightarrow \infty ) \rightarrow 0^{+}$ the function $f(x)$ for $x\geq 0$ is something like Figure 2.
\begin{figure}
  \centering
  \includegraphics[width=1\textwidth]%
    {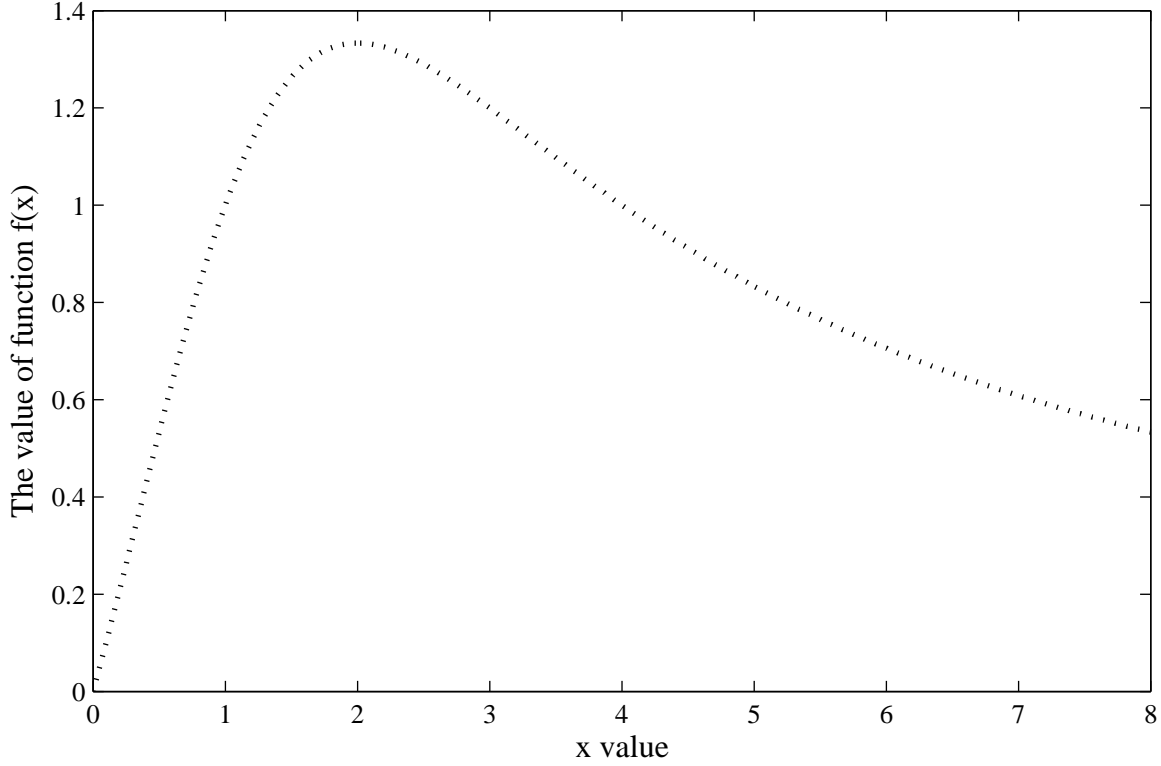}% picture filename
  \caption{The function $f(x)=\frac{Kx}{x^2-x+K}$ versus continuous variable of $x$ for $K=4$.}
\end{figure} 
Therefore, the maximum value of the $d(r)$ can be achieved by finding out the minimum value of $r\in \mathbb{N}$ such that:
\begin{equation}
d(r+1)-d(r)\leq 0.
\end{equation} 
In order to find $r$ to satisfy $d(r+1)-d(r)\leq 0$ condition we have:
\begin{align}
d(r+1)-d(r)&=\frac{K(r+1)}{\underbrace{(r+1)^2-(r+1)+K}_{>0}}-\frac{Kr}{\underbrace{r^2-r+K}_{>0}}&\\
&=\frac{K(r+1)(r^2-r+K)-Kr\left((r+1)^2-(r+1)+K \right)}{\underbrace{\left((r+1)^2-(r+1)+K\right)\left(r^2-r+K\right)}_{>0}}\\
&=\frac{-K\left( r^2+r-K \right)}{\underbrace{\left((r+1)^2-(r+1)+K\right)\left(r^2-r+K\right)}_{>0}} \leq 0&\\
&\Rightarrow r \geq \frac{\sqrt{1+4K}-1}{2},&
\end{align}
Therefore, the minimum value of $r \in \mathbb{N}$ which satisfies above equation is $ r^{*}=\left \lceil{\frac{\sqrt{1+4K}-1}{2}} \right \rceil$.
The exact value of $d(r^{*})$ has been shown for different values of $K$ in Figure 2.
Thus, for a large number of users, the sum DoF of BIA in the $K$-user interference channel approaches $\frac{\sqrt{K}}{2}$. In the following section, we propose an algorithm to systematically generate the antenna switching patterns and the beamforming vectors such that the $\frac{Kr}{r^2-r+K}$ sum DoF is achieved.

\section{Achievable DoF Using Staggered Antenna Switching}

\subsection{Beamforming vectors generation}

 To design beamforming vectors, we assume all the elements of the beamforming vectors are binary, thus ${\bf v}_{d}^{[i]}(j) \in \{0,1\}$. Let's design the precoder matrices and switching pattern from the basic matrix of $\bf{{S}}$. The basic matrix ${\bf{{S}}} \in {\{0,1\}}^{n\times K}$ can be expressed as follows:
\begin{equation}
{\bf {S}}^{\mathrm{T}} = \left[{\underbrace{{\bf A},\dots,{\bf A}}_{\text{r-1 times}}, {\bf B}_{(n-(r-1)K)\times K}}\right] 
\end{equation}
where, $n=\binom{K-1}{r}+r\binom{K-1}{r-1}$, ${\bf A}={\bf 1}_{K \times K}-{\bf I}_{K \times K}$ and ${\bf B}_{(n-(r-1)K)\times K}$ is a matrix with distinct rows and each row containing exactly $K-r$ ones. Also ${\bf 1}_{K \times K}$ is an all-ones square matrix and ${\bf I}_{K \times K}$ is an identity matrix. For instance, in the case of $K = 4$ and $r=3$, the matrix ${\bf S}$ can be represented as follows (take note $r=3$ is not the optimum value for the $K=4$):
\begin{equation}
{\bf {S}}^{\mathrm{T}}=
\left[ \begin{array}{cccccccccccc}
  0&1&1&1&0&1&1&1&1&0&0&0\\
  1&0&1&1&1&0&1&1&0&1&0&0\\
  1&1&0&1&1&1&0&1&0&0&1&0\\
  1&1&1&0&1&1&1&0&0&0&0&1
\end{array} \right]
\end{equation}
 The matrix $\bf S$ consists of $K$ columns where $j^{th}$ column of this matrix is expressed by $\bf {S}_j$. In this case all the basic column vectors of the precoder matrix $\bar{\bf{V}}^{[p]}$ at $\mathrm{TX}^{p}$ is chosen from the following set:
\begin{equation}
\label{preco}
V^{[p]}=\Big{\lbrace} {{{\bf {S}}_{i_1}} \circ {{\bf {S}}_{i_2}} \circ ... \circ {{\bf {S}}_{i_{K-r}}} \Big{\vert} ~i_l \in \{1,\dots,K\}-\{p\}} \Big{\rbrace}.
\end{equation}
It means that all the precoder matrices have the size of $n \times \binom{K-1}{K-r}$ or equivalently have the size of $n \times \binom{K-1}{K-r} = \binom{K-1}{r-1}$.
Thus every $r$ different transmitter like $i_1,~i_2,\dots$ and $i_r$ have exactly one shared basic vector which can be represented as follows:
\begin{equation}
\label{precoderdesign}
\Big{\vert} {\bigcap_{i_1,\dots,i_r}{V^{[i]}}} \Big{\vert} = 1,
\end{equation}
in other words:
\begin{equation}
\label{sharebasicvector}
{\bf v}_{l_{i_1}}^{[i_1]}={\bf v}_{l_{i_2}}^{[i_2]}=\dots={\bf v}_{l_{i_r}}^{[i_r]}={{\bf {S}}_{i'_1}} \circ {{\bf {S}}_{i'_2}} \circ ... \circ {{\bf{S}}_{i'_{K-r}}},~i'_1,i'_2,\dots,i'_{K-r} \in \{1,\dots,K\}-\{i_1,\dots,i_r\}.
\end{equation} 

\subsection{Antenna Switching Pattern at the Receivers}
As it was declared in section II, each receiver equipped with a multi-mode antenna can select among $r$ different receiving paths. Therefore, for the switching pattern ${\bf {SW}}_p={\left[{sw_p (1),\dots,sw_p (n)}\right]}^{\mathrm{T}}$ where $sw_p (j) \in \{0,\dots,M-1\}$ we should find proper ${\bf {SW}}_p$ among $M^{n}$ different switching patterns to satisfy the following conditions:
\begin{itemize}
\item {The shared basic vector ${\bf v}^{[p]}_i= \bigcap_{p \in \{p_1,\dots,p_r\}}{V^{[p]}}$ which is used commonly at $\{{\mathrm{TX}}_{p_1},\dots,{\mathrm{TX}}_{p_r}\}$ after being multiplied by ${\bf{\bar{H}}}^{[lm]}, l \in \{1,\dots,K\}-\{p_1,\dots,p_r\}, m \in \{ p_1,\dots,p_r\}$ should be aligned at their complimentary receivers $\mathrm{RX}_m, m \in \{1,\dots,k\}-\{p_1,\dots,p_r\}$.}
\item {The shared basic vector ${\bf v}^{[p]}_i= \bigcap_{p \in \{p_1,\dots,p_r\}}{V^{[p]}}$ which is used commonly at $\{{\mathrm{TX}}_{p_1},\dots,{\mathrm{TX}}_{p_r}\}$ after being multiplied by ${\bf{\bar {H}}}^{[lm]}, l, m \in \{ p_1,\dots,p_r\}$ channel matrices should be linearly independent of each other at their corresponding receivers $\mathrm{RX}_m, l,m \in \{p_1,\dots,p_r\}$.}
\end{itemize}
Assume that the matrix $\bf {SW}$ is an $n \times K$ matrix which is defined as follows:
\begin{equation}
\label{switchingpattern}
{\bf {SW}}^{\mathrm{T}} = \left[{{\bf {A}},{\bf {A}}+2{\bf I}_{K \times K},\dots,{\bf {A}}+r{\bf I}_{K \times K}, {\bf B}_{(n-(r-1)K)\times K}}\right], 
\end{equation}
Now, let the $\bf{SW}_p$ be the antenna switching pattern at $\mathrm{RX}_p$ which is represented by $p^{th}$ column of the matrix $\bf{SW}$ as follows:
\begin{equation}
{\bf{SW}}_p = {\left[ sw_{1p}, sw_{2p}, \dots ,sw_{np} \right]}^{\mathrm{T}}.
\end{equation}
where $sw_{ip}$ shows $i^{th}$ row and $p^{th}$ column of the matrix $\bf SW$. Therefore, in our designed switching pattern each receiver has been equipped with single antenna by $r$ different receiving mode.
Every basic vector like ${\bf{v}}^{[j]}_{l_j}$ can be equivalently expressed by $r+1$ sub-matrices as follows:
\begin{equation}
\label{vectorsubmatrix}
{\bf{v}}^{[q]}_{i}={\left[{\left({\bf{v}}^{[q]}_{{{\bf e_1} i}}\right)^{\mathrm{T}}},\dots,{\left({\bf{v}}^{[q]}_{{{\bf e_r} i}}\right)^{\mathrm{T}}},{\left({\bf{v}}^{[q]}_{{{\bf f} i}}\right)^{\mathrm{T}}}\right]}^{\mathrm{T}}
\end{equation}
where ${\bf{v}}^{[q]}_{{{\bf e_j} i}}, j \in \{1,\dots,r\}$ and ${\bf{v}}^{[q]}_{{{\bf f} i}}$ are two $K \times 1$ and $\left({n-(r-1)K}\right) \times 1$ column matrices, respectively. Now we must show that all the basic vectors generated at the specific transmitter like $\mathrm{TX}_q$ are linearly independent. The following lemma shows that every generated basic vector from \eqref{precoderdesign} are linearly independent.
\begin{lem}
{\it For all the values of the number of the users $K$ and $r=\lceil{\frac{\sqrt{1+4K}-1}{2}}\rceil$, every generated basic vector at a specific transmitter e.g. $\mathrm{TX}_j$ from Hadamard product of all the combination of $K-r$ column vectors of the ${\bf {S}}$ are linearly independent.}
\end{lem}
\begin{IEEEproof}
Consider $\mathrm{TX}_q$, all the basic vectors of this transmitter chosen from the following set:
\begin{equation}
V^{[p]}=\Big{\lbrace} {{{\bf {S}}_{i_1}} \circ {{\bf {S}}_{i_2}} \circ ... \circ {{\bf {S}}_{i_{K-r}}} \Big{\vert} ~i_l \in \{1,\dots,K\}-\{p\}} \Big{\rbrace}.
\end{equation} 
We must show that at $\mathrm{TX}_{p}$ where $\bar{{\bf{V}}}^{[p]}=\left[{{\bf {v}}^{[p]}_{1},\dots,{\bf {v}}^{[p]}_{\binom{K-1}{r-1}}}\right] $, all the sub-matrices ${\bf{v}}^{[p]}_{f i_l}=\left[{{v}}^{[p]}_{f i_l}(1),\dots,{{v}}^{[p]}_{f i_l}(n-(r-1)K)\right], i_l \in \{1,\dots,\binom{K-1}{r-1}\}$ are linearly independent. Since the matrix ${\bf{B}}_{(n-(r-1)K)\times K}$ generates the elements of these vectors we have to analyze it. Each row of the matrix ${\bf{B}}_{(n-(r-1)K)\times K}$ contains exactly $K-r$ ones. If ${\bf{B}}_{(n-(r-1)K)\times K}=[b_{ij}], 1\leq i \leq (n-(r-1)K),~ 1 \leq j \leq K$, it means that for generating nonzero element e.g. ${{v}}^{[p]}_{f i_l}(u)$ in the especial position of the sub-matrix ${\bf {v}}^{[p]}_{f i_l}, 1 \leq i_l \leq \binom{K-1}{r-1}$ which is shared among $\{p^{th}_1,\dots,p^{th}_r \}$ transmitters we must have:
\begin{equation}
\label{po}
b_{((r-1)K+u)p'_1} ~b_{((r-1)K+u)p'_2}~ \dots~b_{((r-1)K+u)p'_{(K-r)}} \neq 0,~p'_1,\dots,p'_{K-r} \in \{1,\dots,K\}-\{p_1,\dots,p_r \}
\end{equation}
In appendix I we show that for every value of $K$ and $r=\lceil{\frac{\sqrt{1+4K}-1}{2}}\rceil$ the value of $\left(n-(r-1)K\right)-\binom{K}{r} \geq -1$. For the case of $\left(n-(r-1)K\right)-\binom{K}{r} = -1$ except one of ${\bf{v}}^{[p]}_{f i_l}, 1 \leq i_l \leq \binom{K-1}{r-1}$ which is all zero matrix, referring to equation \eqref{po}, all the sub-matrices ${\bf{v}}^{[p]}_{f i_l}$ have a nonzero element in the unique position. For the case of $\left(n-(r-1)K\right)-\binom{K}{r} \geq 0$, it is clear from \eqref{po} that all the sub-matrices ${\bf{v}}^{[p]}_{f i_l}$ have at least one nonzero element in the unique position. Therefore, for all values of $K$ and $r=\lceil{\frac{\sqrt{1+4K}-1}{2}}\rceil$ all the generated ${\bf{v}}^{[p]}_{f i_l}, 1 \leq i_l \leq \binom{K-1}{r-1}$ are linearly independent. Since ${\bf{v}}^{[p]}_{f i_l}, 1 \leq i_l \leq \binom{K-1}{r-1}$ are the sub-matrices of the basic vectors of ${\bf v}^{[p]}_{i_l}, 1 \leq i_l \leq \binom{K-1}{r-1}$ are linearly independent too. Therefore, the proof was completed.  
%\IEEEQEDhere
\end{IEEEproof}
\begin{lem}
{\it Using ${\bf SW}_l$ at $\mathrm{RX}_l$ for every basic vector ${\bf v}^{[m]}_i=\bigcap _{p \in \{p_1,\dots,p_r\}}{V^{[p]}}$, the received vectors $\bar{{\bf H}}^{[lm]}{\bf v}^{[m]}_i, m \in \{{p_1,p_2,\dots,p_r}\}, l \in \{1,\dots,K\}-\{p_1,\dots,p_r\}$ are aligned with each other.}
\end{lem}
\begin{IEEEproof}
The proof was provided by analyzing both nonzero elements of the basic vector ${\bf v}^{[m]}_i$ and the diagonal matrix of $\bar{\bf H}^{[lm]}$. Similar to \eqref{vectorsubmatrix}, the basic vector of ${\bf v}^{[m]}_i$ can be represented by the $r$ sub-matrices as follows:
\begin{equation}
\label{vectorrep}
{\bf{v}}^{[m]}_{i}={\left[{\left({\bf{v}}^{[m]}_{{{\bf e} i}}\right)^{\mathrm{T}}},\dots,{\left({\bf{v}}^{[m]}_{{{\bf e} i}}\right)^{\mathrm{T}}},{\left({\bf{v}}^{[m]}_{{{\bf f} i}}\right)^{\mathrm{T}}}\right]}^{\mathrm{T}}.
\end{equation} 
From \eqref{sharebasicvector}, for the matrix ${\bf{v}}^{[m]}_{{{\bf e} i}}=\left[{v^{[m]}_{{{\bf e} i}}}(1),{v^{[m]}_{{{\bf e} i}}}(2),\dots,{v^{[m]}_{{{\bf e} i}}}(K)\right]$ we have:
\begin{equation}
{v^{[m]}_{{{\bf e} i}}}({p_1})={v^{[m]}_{{{\bf e} i}}}({p_2})=\dots={v^{[m]}_{{{\bf e} i}}}({p_r})=1.
\end{equation}
It means that the only nonzero elements of ${\bf{v}}^{[m]}_{{{\bf e} i}}$ is its $\{ {p_1}^{th},{p_2}^{th},\dots,{p_r}^{th} \}$ elements. Similarly for the switching pattern at $l^{th}$ receiver e.g. ${\bf{SW}}_l \left[{1:(r-1)K}\right]={\left[{sw_{l}(1),\dots,{sw_{l}((r-1)K)}}\right]}^{\mathrm{T}}$ we have:
\begin{equation}
sw_l(i)=1, i\in \lbrace{p_1,\dots,p_r,\dots,(r-2)K+p_1,\dots,(r-2)K+p_r}\rbrace,
\end{equation}
similarly for nonzero elements ${\bf{v}}^{[m]}_{{{\bf f} i}}=\left[{{v}^{[m]}_{{\bf f} i}(1)},\dots,{{v}^{[m]}_{{\bf f} i}(n-(r-1)K)}\right]^{\mathrm{T}}$ e.g. ${{v}^{[m]}_{{\bf f} i}(j)}=1$ the value of $sw_l(i+(r-1)K)$ is equal to 1. Therefore, at $\mathrm{RX}_l, l \in \{1,\dots,K\}-\{p_1,\dots,p_r\}$ all the basic vectors like ${\bf{v}}^{[m]}_{i}, m \in \{p_1,\dots,p_r\}$ received by multiplying by the constant number of $h^{[lm]}\left(1\right)$. Thus all the ${\bar{\bf{H}}}^{[lm]}{\bf v}^{[m]}_i$, $m \in \{p_1,\dots,p_r\}$ and $l \in \{1,\dots,K\}-\{p_1,\dots,p_r\}$ arrive along  the basic vector of ${\bf v}^{[m]}_i$. So the proof is completed.
%\IEEEQEDhere
\end{IEEEproof}
\begin{lem}
{\it Using ${\bf SW}_l$ at $\mathrm{RX}_l$ for every basic vector ${\bf v}^{[m]}_i=\bigcap _{p \in \{p_1,\dots,p_r\}}{V^{[p]}}$, the received vectors $\bar{{\bf H}}^{[lm]}{\bf v}^{[m]}_i, l,m \in \{{p_1,p_2,\dots,p_r}\}$ are linearly independent.}
\end{lem}
\begin{IEEEproof}
The basic vector ${\bf{v}}^{[m]}_{i}$ similar to \eqref{vectorrep} can be represented by the following equation:
\begin{equation}
{\bf{v}}^{[m]}_{i}={\left[{\left({\bf{v}}^{[m]}_{{{\bf e} i}}\right)^{\mathrm{T}}},\dots,{\left({\bf{v}}^{[m]}_{{{\bf e} i}}\right)^{\mathrm{T}}},{\left({\bf{v}}^{[m]}_{{{\bf f} i}}\right)^{\mathrm{T}}}\right]}^{\mathrm{T}}.
\end{equation}
If we show that at $\mathrm{RX}_m , m \in \{p_1,\dots,p_r\}$, all the $\bar{{\bf H}}^{[lm]}(1:(r-1)K){\left[{\left({\bf{v}}^{[m]}_{{{\bf e} i}}\right)^{\mathrm{T}}},\dots,{\left({\bf{v}}^{[m]}_{{{\bf e} i}}\right)^{\mathrm{T}}}\right]}^{\mathrm{T}}$ and $l,m \in \{p_1,\dots,p_r\}$ are linearly independent, the proof will be accomplished. Since all the nonzero elements of ${\left[{\left({\bf{v}}^{[m]}_{{{\bf e} i}}\right)^{\mathrm{T}}},\dots,{\left({\bf{v}}^{[m]}_{{{\bf e} i}}\right)^{\mathrm{T}}}\right]}^{\mathrm{T}}$ are in  the set of $\lbrace{p_1,\dots,p_r,\dots,(r-2)K+p_1,\dots,(r-2)K+p_r}\rbrace$ and from \eqref{switchingpattern} all the first $(r-1)K$ elements of the channels have the following form:
\begin{equation}
\begin{aligned}
&{\bar{\bf{H}}}^{[lm]}{\left(1:(r-1)K\right)}=&\\
&\mathrm{diag}{\left[{h^{[lm]}_1(1)},\dots,{h^{[lm]}_{q_1}(0)},\dots,{h^{[lm]}_{K}(0)},\dots,{h^{[lm]}_{K+q_1}(2)},\dots,{h^{[lm]}_{q_1+(r-2)K}(r-1)},\dots,{h^{[lm]}_{(r-1)K}(1)}\right]},&
\end{aligned}
\end{equation}
the common received basic vectors from $\mathrm{TX}_p, p\in \{p_1,\dots,p_r\}$ at $\mathrm{RX}_l, l \in \{p_1,\dots,p_r\}$ at least have $r$ different elements. Therefore, all the ${\bar{\bf{H}}}^{[lm]}{\left(1:(r-1)K\right)} {\bf v}^{[m]}_l \left(1:(r-1)K\right)$, $m \in \{p_1,\dots,p_r\}$ are linearly independent. So the proof is completed.
%\IEEEQEDhere
\end{IEEEproof}
 In the next section, we show that using the designed switching antenna pattern and the designed precoders, the $\frac{Kr}{r^2-r+K}$ sum DoF can be achieved. Using designed switching pattern assumptions has important hardware implications. For instance, the proposed algorithm operates with low cost reconfigurable antennas that have only $r$ modes. Besides, beamforming is very simple and applied by activating or deactivating certain symbols at the transmitter.

\subsection{DoF achievability using proposed switching pattern and the designed precoders}
Now we want to show that by the designed precoders the DoF of $\frac{Kr}{r^2-r+K}, r=\lceil{\frac{\sqrt{1+4K}-1}{2}}\rceil$ is achievable. In our designed precoders every transmitter e.g. $\mathrm{TX}_j$ has $\binom{K-1}{r-1}$ basic vectors. From Lemma1 every generated basic vector at $\mathrm{TX}_j$ are linearly independent. Therefore, the total dimension used at each transmitter is equal to $\binom{K-1}{r-1}$. Now we must show that these basic vectors are linearly independent at their corresponding receivers. In order to analyze this fact we should show that these basic vectors are linearly independent from basic vectors of other transmitters at $\mathrm{RX}_j$. The generated basic vector at each transmitter has two different types as follows:
\begin{enumerate}
\item The basic vectors which are linearly independent from each other.
\item The basic vectors which are shared among other transmitters.
\end{enumerate}
The type one vectors which are linearly independent  from each other, because of $\lvert{h_j^{[pq]}}\rvert > 0,~p,q \in\{1,\dots,K\}$ also remain linearly independent at their receivers. From the point of view of the $\mathrm{RX}_j$ and from Lemma2, the basic vectors which are not shared with the basic vectors of the $\mathrm{TX}_j$ are aligned with each others. The number of such basic vectors can be calculated by the number of countable $r$ different transmitters among $K-1$ transmitters (except $\mathrm{TX}_j$). Therefore, the number of such vectors is $\binom{K-1}{r}$. Also, there is some basic vectors which are shared among $j^{th}$ transmitter and all other transmitters. The number of such vectors can be calculated by counting the number of $r-1$ choosable transmitters among $K-1$ ones which is equal to $\binom{K-1}{r-1}$. From Lemma3 such vectors are linearly independent and therefore occupy $r\binom{K-1}{r-1}$ dimensions at $j^{th}$ receiver. Therefore, at $\mathrm{RX}_j$ the $\mathrm{TX}_j$ occupies $\binom{K-1}{r-1}$ dimensions (desired signal space dimensions) at its corresponding receiver. Also, we have $\left({r-1}\right)\binom{K-1}{r-1}$ dimensions which are generated by the basic vectors shared among $\mathrm{TX}_j$ and all other transmitters. These basic vectors from Lemma 3 are linearly independent and the total dimensions occupied by such vectors is $(r-1)\binom{K-1}{r-1}+\binom{K-1}{r-1}=r\binom{K-1}{r-1}$. On the other hand, at $\mathrm{RX}_j$ there are some basic vectors which are not shared by $j^{th}$ transmitter. The number of such basic vectors can be calculated by counting $r$ different choosable transmitters among $K-1$ ones (except $\mathrm{TX}_j$) which is equal to $\binom{K-1}{r}$. Therefore the total number of dimensions is equal to summing $r\binom{K-1}{r-1}$ and $\binom{K-1}{r}$ dimensions which is equal to $r\binom{K-1}{r-1}+\binom{K-1}{r}$. The number of desired signal dimensions at $\mathrm{RX}_j$ is equal to $\binom{K-1}{r-1}$, which means that the total number of desired signal dimensions at $j^{th}$ user equals to $\binom{K-1}{r-1}$ from $r\binom{K-1}{r-1}+\binom{K-1}{r}$ transmission time slots and consequently the DoF of $\frac{\binom{K-1}{r-1}}{r\binom{K-1}{r-1}+\binom{K-1}{r}}=\frac{r}{r^2-r+K}$ for $j^{th}$ user can be achievable. By the similar method of proof we can show that all other transmitters can get to $\frac{r}{r^2-r+K}$ DoF and the $K-$user interference network totally can reach to the $\frac{Kr}{r^2-r+K}$ DoF, which meets the upper-bound. Figure 4 shows DoF rate region of $K-$user interference channel using staggered antenna switching. The result shows that the proposed method in \cite{Alaa} traces our method for $2 \leq K \leq 6$ and satisfies the rate region proposed by Wang in \cite{7}.   
\begin{figure}
  \centering
  \includegraphics[width=1\textwidth]%
    {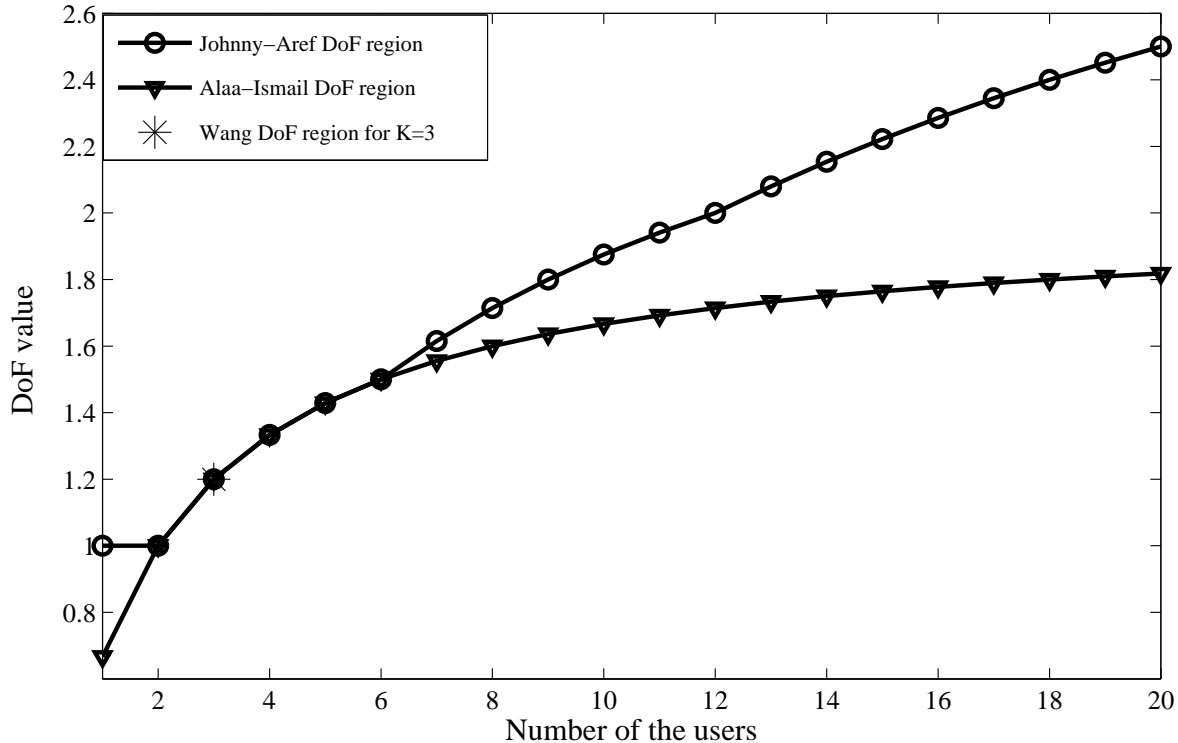}% picture filename
  \caption{DoF rate region of $K-$user interference channel versus different number of the users $K$.}
\end{figure}      
\subsection{5-user SISO Interference Channel alignment using staggered antenna switching}
Consider a fully connected 5-user SISO Interference Channel. The maximum achievable DoF in this case can be found by setting $r=\left \lceil{\frac{\sqrt{1+4K}-1}{2}} \right \rceil{\vert_{K=5}}=2$. In this setting every transmitter can send 4 symbols through 14 time slots. In order to design precoders first of all we demonstrate the matrix $\bf S$ as follows:
\begin{equation}
\label{15}
{\bf {S}}^{\mathrm{T}}=
\left[ \begin{array}{cccccccccccccc}
         0 & 1 & 1 & 1 & 1 & 0 & 0 & 0 & 0 & 1 & 1 & 1 & 1 & 1\\
		 1 & 0 & 1 & 1 & 1 & 0 & 1 & 1 & 1 & 0 & 0 & 0 & 1 & 1\\
		 1 & 1 & 0 & 1 & 1 & 1 & 0 & 1 & 1 & 0 & 1 & 1 & 0 & 0\\
		 1 & 1 & 1 & 0 & 1 & 1 & 1 & 0 & 1 & 1 & 0 & 1 & 0 & 1\\
		 1 & 1 & 1 & 1 & 0 & 1 & 1 & 1 & 0 & 1 & 1 & 0 & 1 & 0
\end{array} \right]
\end{equation}	
In this case since $r=2$, the value of the matrix $\bf SW=\bf S$. Also, from \eqref{preco}, we can design all the $\binom{5}{2}=10$ basic vectors at each transmitter as follows:
\begin{align}
&{\bf v}_1^{[1]}={\bf v}_1^{[2]}={\left[ {1~1~0~0~0~1~0~0~0~0~0~0~0~0}\right]}^{\mathrm{T}}&\\
&{\bf v}_2^{[1]}={\bf v}_1^{[3]}={\left[ {1~0~1~0~0~0~1~0~0~0~0~0~0~0}\right]}^{\mathrm{T}}&\\
&{\bf v}_3^{[1]}={\bf v}_1^{[4]}={\left[ {1~0~0~1~0~0~0~1~0~0~0~0~0~0}\right]}^{\mathrm{T}}&\\
&{\bf v}_4^{[1]}={\bf v}_1^{[5]}={\left[ {1~0~0~0~1~0~0~0~1~0~0~0~0~0}\right]}^{\mathrm{T}}&\\
&{\bf v}_2^{[2]}={\bf v}_2^{[3]}={\left[ {0~1~1~0~0~0~0~0~0~1~0~0~0~0}\right]}^{\mathrm{T}}&\\
&{\bf v}_3^{[2]}={\bf v}_2^{[4]}={\left[ {0~1~0~1~0~0~0~0~0~0~1~0~0~0}\right]}^{\mathrm{T}}&\\
&{\bf v}_4^{[2]}={\bf v}_2^{[5]}={\left[ {0~1~0~0~1~0~0~0~0~0~0~1~0~0}\right]}^{\mathrm{T}}&\\
&{\bf v}_3^{[3]}={\bf v}_3^{[4]}={\left[ {0~0~1~1~0~0~0~0~0~0~0~0~1~0}\right]}^{\mathrm{T}}&\\
&{\bf v}_4^{[3]}={\bf v}_3^{[5]}={\left[ {0~0~1~0~1~0~0~0~0~0~0~0~0~1}\right]}^{\mathrm{T}}&\\
&{\bf v}_4^{[4]}={\bf v}_4^{[5]}={\left[ {0~0~0~1~1~0~0~0~0~0~0~0~0~0}\right]}^{\mathrm{T}}.&
\end{align} 
As it was proved in Lemma 1, all the generated basic vectors at each transmitter are linearly independent e.g. ${\bf v}_1^{[1]}$, ${\bf v}_2^{[1]}$, ${\bf v}_3^{[1]}$ and ${\bf v}_4^{[1]}$. Now we can design the switching pattern at each receiver. In this case since the optimum value of $r$ is equal to 2, every receiver is equipped with an antenna with two switching modes. Therefore, each receiver during data reception can switch between its two reception paths. From \eqref{switchingpattern} we can get switching pattern at each receiver as follows:
\begin{align}
&{\bf SW}_1=\left[0~1~1~1~1~0~0~0~0~1~1~1~1~1\right]^{\mathrm{T}}&\\
&{\bf SW}_2=\left[1~0~1~1~1~0~1~1~1~0~0~0~1~1\right]^{\mathrm{T}}&\\
&{\bf SW}_3=\left[1~1~0~1~1~1~0~1~1~0~1~1~0~0\right]^{\mathrm{T}}&\\
&{\bf SW}_4=\left[1~1~1~0~1~1~1~0~1~1~0~1~0~1\right]^{\mathrm{T}}&\\
&{\bf SW}_5=\left[1~1~1~1~0~1~1~1~0~1~1~0~1~0\right]^{\mathrm{T}}.&
\end{align} 
In this case due to the above switching pattern, for receiver one, we have the following channel form:
\begin{equation}
{\bar {\bf H}}^{[1q]}=\mathrm{diag}\left({\left[{h_1^{[1q]}(0),h_2^{[1q]}(1),\dots,h_5^{[1q]}(1),h_6^{[1q]}(0),\dots,h_9^{[1q]}(0),h_{10}^{[1q]}(1),\dots,h_{14}^{[1q]}(1)}\right]}\right)
\end{equation}
Therefore, the members of the following set ($S^{[1]}$), show the basic vectors which span the space of the first receiver:
\begin{equation}
\begin{aligned}
S^{[1]}=& \{\underbrace{{\bar{\bf H}}^{[11]}{\bf v}_1^{[1]},{\bar{\bf H}}^{[12]}{\bf v}_1^{[2]}}_{\text{linearly independent}},\underbrace{{\bar{\bf H}}^{[11]}{\bf v}_2^{[1]},{\bar{\bf H}}^{[13]}{\bf v}_1^{[3]}}_{\text{linearly independent}},\underbrace{{\bar{\bf H}}^{[11]}{\bf v}_3^{[1]},{\bar{\bf H}}^{[14]}{\bf v}_1^{[4]}}_{\text{linearly independent}},\underbrace{{\bar{\bf H}}^{[11]}{\bf v}_4^{[1]},{\bar{\bf H}}^{[12]}{\bf v}_1^{[5]}}_{\text{linearly independent}}&\\
&\underbrace{{\bar{\bf H}}^{[12]}{\bf v}_2^{[2]},{\bar{\bf H}}^{[13]}{\bf v}_2^{[3]}}_{\text{align}},\underbrace{{\bar{\bf H}}^{[12]}{\bf v}_3^{[2]},{\bar{\bf H}}^{[13]}{\bf v}_2^{[4]}}_{\text{align}},\underbrace{{\bar{\bf H}}^{[11]}{\bf v}_4^{[2]},{\bar{\bf H}}^{[15]}{\bf v}_2^{[5]}}_{\text{align}},\underbrace{{\bar{\bf H}}^{[12]}{\bf v}_3^{[3]},{\bar{\bf H}}^{[12]}{\bf v}_3^{[4]}}_{\text{align}}&\\
&\underbrace{{\bar{\bf H}}^{[12]}{\bf v}_4^{[3]},{\bar{\bf H}}^{[13]}{\bf v}_3^{[5]}}_{\text{align}},\underbrace{{\bar{\bf H}}^{[12]}{\bf v}_4^{[4]},{\bar{\bf H}}^{[13]}{\bf v}_4^{[5]}}_{\text{align}} \}.&
\end{aligned}
\end{equation} 
Since $\bar{\bf H}^{[1q]}$, in the time slots of $\{2,3,4,5,10,11,12,14\}$ and $\{1,6,7,8,9\}$ experiences similar coefficients of $h^{[1q]}(1)$ and $h^{[1q]}(0)$ respectively, the basic vectors of ${\bf v}^{[j]}_i, i>1, j \neq 1$ are aligned with ${\bar{\bf H}}^{[1j]}{\bf v}^{[j]}_i, i>1, j \neq 1$. In other words, in this case we have:
\begin{equation}
\mathrm{dim}\left({\left[ {{\bar{\bf H}}^{[1j]}{\bf v}^{[j]}_i~{\bf v}^{[j]}_i} \right] }\right)=1,~i>1, j \neq 1.
\end{equation}  	
The above relation shows that all the shared generated basic vectors such as $\{{\bf v}_2^{[2]},{\bf v}_2^{[3]}\}$, $\{{\bf v}_3^{[2]},{\bf v}_2^{[4]}\}$, $\{{\bf v}_4^{[2]},{\bf v}_2^{[5]}\}$, $\{{\bf v}_3^{[3]},{\bf v}_3^{[4]}\}$, $\{{\bf v}_4^{[3]},{\bf v}_3^{[5]}\}$ and $\{{\bf v}_4^{[4]},{\bf v}_4^{[5]}\}$ after being multiplied by channel matrices of ${\bar{\bf H}}^{[1j]}, i>1, j \neq 1$ remain aligned with each other. In this case since the basic vectors of $\{{\bf v}_1^{[1]},{\bf v}_1^{[2]}\}$ have the nonzero elements in the time slots of $\{1,2,6\}$ and the channel model matrix changes its value between time slots of one and two, both ${\bar{\bf H}}^{[11]}{\bf v}_1^{[1]}$ and ${\bar{\bf H}}^{[12]}{\bf v}_1^{[2]}$ are linearly independent. Similarly all other received basic vectors of $\{\bar{\bf H}^{[11]}{\bf v}_2^{[1]},\bar{\bf H}^{[13]}{\bf v}_1^{[3]}\}$, $\{\bar{\bf H}^{[11]}{\bf v}_3^{[1]},\bar{\bf H}^{[14]}{\bf v}_1^{[4]}\}$ and $\{\bar{\bf H}^{[11]}{\bf v}_4^{[1]},\bar{\bf H}^{[15]}{\bf v}_1^{[5]}\}$ are jointly linearly independent. Therefore, at the first receiver from 14 dimensions we have four free interference dimensions and this user can achieve $\frac{4}{14}$ DoF. Similarly we can achieve $\frac{4}{14}$ for all other users and totally we get $\frac{10}{7}$ DoF. 

\section{Conclusion}
In this paper, we have shown that in the $K$-user SISO interference channel the DoF region of linear Blind Interference Alignment (BIA) using staggered antenna switching is $\max_{r \in \mathbb{N}}\frac{Kr}{r^2-r+K}$. We show both achievability and converse proof for this important problem. A key insight is that each signal dimension from one user can be aligned with a set of distinct users at the receivers of the rest of the users.  Without channel state information at the transmitters, this result indicates that when the value of $K$ limits to infinity we can achieve $\frac{\sqrt{K}}{2}$ compared to the unity achievable DoF of the orthogonal multiple access schemes. Moreover, we proposed an algorithm to generate the transmit beamforming vectors and antenna switching patterns utilized in BIA. We showed that the proposed algorithm can achieve the $\frac{Kr}{r^2-r+K}$ sum DoF for any $K$ and $r$ values. Also we show that the term $\frac{Kr}{r^2-r+K}$ is maximized when the value of $r \in \mathbb{N}$ is equal to $\left \lceil{\frac{\sqrt{1+4K}-1}{2}} \right \rceil$. By applying both achievable method and converse proof of this work for the 3-user Interference Channel, we showed that a sum DoF of $\frac{6}{5}$, which was obtained previously in \cite{7} was met. 
\appendices
\section{Proof of the inequality}
\begin{lem}{\it For all values of $K$ and $r=\lceil {\frac{\sqrt{1+4K}-1}{2}} \rceil$, we can conclude that $r \binom{K-1}{r-1}+\binom{K-1}{r}-1 \geq (r-1)K+\binom{K}{r}-1$.}
\end{lem}
\begin{IEEEproof}
Starting from finding the sign of the term $r \binom{K-1}{r-1}+\binom{K-1}{r}-(r-1)K-\binom{K}{r}+1$, we have:
\begin{align}
\mathrm{sgn} &\left(r \binom{K-1}{r-1}+\binom{K-1}{r}-(r-1)K-\binom{K}{r}+1\right )&\\
&=\mathrm{sgn}\left({(r-1)\binom{K-1}{r-1}+\binom{K-1}{r-1}+\binom{K-1}{r}-\binom{K}{r}}-(r-1)K+1\right)&\\
&=\mathrm{sgn}\left({(r-1)\left({\binom{K-1}{r-1}-K}\right)+1}\right)&
\end{align}
For $K \leq 6$ where, $r=2$ the term ${(r-1)\left({\binom{K-1}{r-1}-K}\right)+1}$ can be simplified as follows:
\begin{equation}
{\left({\binom{K-1}{1}-K}\right)+1}=0,~K \leq 6,
\end{equation}
which satisfies the result of this lemma. For $K > 6$, by the use of Stirling's approximation, we can easily show that the term $\binom{K-1}{r-1}$ is strictly larger than the value of $K$. Therefore, the term ${\binom{K-1}{r-1}-K}$ is surely larger than zero and the proof of this lemma is completed.
\end{IEEEproof}
\ifCLASSOPTIONcaptionsoff
  \newpage
\fi
%\IEEEtriggeratref{6}

\end{document}